\documentclass{article}
\usepackage{graphicx}
\usepackage{amsmath}

\begin{document}
\markboth{Sanjay Sarwe and R. V. Saraykar}
{Stability Analysis in 
 N dimensional Gravitational collapse with equation of state} 
\title{\bf{ \textrm{Stability Analysis in N Dimensional Gravitational 
 Collapse with Equation of  State}} }
 
\author {{\bf Sanjay Sarwe } \footnote{email:sbsarwe@gmal.com}\\
   Department of Mathematics, S. F. S. College,
  \\ Seminary Hill,
   Nagpur-440 006, India  \\
 {\bf R. V. Saraykar}  \footnote{email:ravindra.saraykar@gmail.com} \\
 Department of Mathematics, Nagpur University \\
 Campus, Nagpur-440 033, India}
\date{}
\maketitle

 \begin{abstract}
 {We study stability of occurrence of black holes and 
   naked singularities that arise as the final states of a complete 
   gravitational collapse of type I matter field in a spherically 
   symmetric $N$ dimensional spacetime with equation of state $p = k \rho$, 
   $0 \leq k \leq 1$.  We prove 
   that for a regular initial data comprising of pressure (or density ) profiles at an
		initial surface $t = t_{i}$, from which the collapse 
   evolves, there exists a large class of  the velocity functions
   and classes of solutions of Einstein equations, such that the spacetime evolution goes 
		to a final state  which is either a black hole or a naked singularity. 
   We use suitable function spaces for regular initial data leading the collapse to a black hole 
	or a naked singularity and show that the data forms an open subset of the set of all regular initial data. 
	In this sense, both the outcomes of collapse are stable.  These results are 
   discussed and analyzed in the light of the cosmic censorship hypothesis
   in black hole physics. }
\end{abstract}   
 
 {\small PAC numbers: 04.20 Dw, 04.20. Jb, 04.70. Bw }  
\section {Introduction}
The arrival of
Tolman's dust solutions
 \cite {Dtol}
  in 1934  to Einstein's field
equations in spherically symmetric spacetime has led to
investigation of nature of singularity in various models like
dust, null dust, perfect fluid, quark matter etc. in later years.
Considering Tolman solution, Oppenheimer and Snyder
 \cite {Irohs},
in 1939, have studied the collapse of spherically symmetric
homogeneous dust cloud, which led to the general concept of
trapped surfaces and the black hole (BH). In 1951, Vaidya
identified the general metric appropriate to a spherically
symmetric distribution of null dust
 \cite {Dvai} 
 and considering
linear mass function, Papapetrou \cite{Dpaku} and Kurada
 \cite{YK} have
shown that the central singularity will be a naked singularity and
persistent, when the collapse is sufficiently slow. Hereafter,
many numerical and theoretical studies were carried out in
different models by various authors
 \cite {Idmels}-\cite{OP}
presenting counter examples to Cosmic Censorship Cenjecture (CCC).

In a recent paper, R. Goswami and P. S. Joshi
 \cite {Drgpj-1}
pointed out that the occurrence of naked singularity (NS) as end state
of dust collapse beginning from regular initial data with
physically relevant assumptions, can be removed when one goes to
higher dimensional space-time. This has restored CCC in dust
collapse for higher dimensions of the space-time. Many
other cases of the dust collapse have been studied by various
authors
\cite{Bgb}-\cite{PGS}.

 We consider Type I matter field since it includes most of the  
physically reasonable forms of matter field, like dust, perfect fluid, etc. The
authors in 
 \cite{Ajd}-\cite{PJ-02}
  have discussed the role of initial
data in spherically symmetric gravitational collapse for Type I
matter fields. 

Goswami and Joshi
\cite{Arg2}
have studied 
the case of an isentropic perfect fluid with linear
equation of state in four dimensional spacetime, wherein, they
pointed out that the occurrence of NS/BH evolving from regular
initial data depends on the choice of rest of the free functions
available.
A. Mahajan et. al. 
   have proved that in $N$ dimensional 
dust collapse with tangential pressure present, the occurrence of NS can be removed when one goes to
higher dimensional spacetime, thus restoring CCC 
\cite{rgpj-1}.

In recent years, special attention is given to higher dimensional
gravitational collapse since the development in string theory
suggest that gravity may be a higher dimensional interaction
\cite{RS01}-\cite{RS-02}.
Hence, it would be interesting to know the status of CCC in higher
dimensional spacetime. These studies may help in possible
appropriate mathematical formulation of the conjecture or assist
in providing any possible proof of CCC
{\cite{Bkk}}-\cite{GSS}. 

 N. Dadhich et.al.
 \cite{DSD}
  have studied spherically symmetric collapse of  fluid with non-vanishing radial pressure in higher dimensional space-time with equation of state 
 $p_{r} =k \rho$ through construction of the root equation governing the nature
(BH versus NS) of the central singularity. 

However, all of the initial data
is not independent, so it poses a question, which initial data set
shall evolve into a NS as a result of gravitational collapse?
Further, having a NS for certain initial data set is not enough
because if it looses its characteristic for small perturbation in
the neighbourhood of that data set then the NS is
not serious enough to challenge cosmic censorship conjecture
 \cite{rp}.
 Therefore, the all important question is that, ``{\it{is the NS
developed from certain initial data set a stable one}}?''. We study 
this
behaviour of NS using characterization of stability with respect to initial data following
Abraham and Marsden
 \cite{Irajm}-\cite{Isbsrvs} .
 
It is relevant to introduce equation of state as it makes the
collapsing models physically more relevant and crystallizes the methodology used in the analysis to bring forth the outcomes. Does the linear
equation of state $p = k \rho$ stimulates the outcome of NS/BH
phases as end state of gravitational collapse? We investigate this
question through analysis of field equations of N-dimensional gravitational
collapse with linear equation of state in section 2. 
The study of radial null geodesics in $N$ dimensions is carried out in section 3.
In section 4, we
analyze the condition of existence of naked singularity to prove
our assertion that the NS
developed from certain initial data set is a stable one by using existence theory of first
order ordinary differential equations
 \cite{Ave}.
  Thereafter, in section 5,
we prove openness of initial data set  ensuring the stability of occurrence
of NS with respect to initial data.
Finally, we comment on genericity of this set.
 
\section{Collapse with linear equation of state }
The general spherically symmetric metric describing $N$ dimensional
 space-time geometry of a collapsing cloud can be described in
 comoving coordinates \\ $(t, r, \theta^{1}, \theta^{2},..,\theta^{N-2})$ by
\begin{equation}
 ds^2 = - e^{2 \nu(t,r)}dt^2 +  e^{2 \psi(t,r)}dr^2 + R^2(t,r) d
\Omega^2_{N-2} \label {Dm01}
\end{equation}
where
\begin{eqnarray*}
 d\Omega^2_{N-2}=\sum_{i=1}^{N-2} \left[\prod_{j=1}^{i-1}
 \sin^2(\theta^{j})\right]\;(d\theta^{i})^2 \nonumber
\end{eqnarray*}
 is the
metric on an $(N-2)$ sphere.
The stress-energy tensor for Type I field
in a diagonal form is given by 
\cite {Ahe}
\begin{equation}
 T{^{t}_{t} } = -\rho, \; T{^{r}_{r} } =
 T{^{\theta^{i}}_{\theta^{i}}} = \; p  \label {Dsem03}
\end{equation}
where $\rho$ and $p$ are the energy
density and pressure respectively. The matter field 
satisfies weak energy condition that implies ${{\rho}{\geq}}0$;
${\rho+p{\geq}} 0$. Linear equation of state for perfect fluid is
\begin{equation}
p(t,r) = k \ \rho(t,r) \ \text{where} \  k \in [0, 1] \; \; . \label
{Des04}
\end{equation}
 Einstein field equations for the metric (\ref {Dm01}) are derived as
  \cite{Arg2}
\begin{equation}
\rho = \frac{(N-2){F}'}{2 R^{N-2}R'} \;  \;
 = - \frac{1} {k} \frac{(N-2)\dot{{F}}}{2 R^{N-2} \dot{R}} \label {Dsfe05}
 \end{equation}
\begin{equation}
\nu' = - \frac{k}{k + 1} [ \ln(\rho)]' \label {Dsfe06}
\end{equation}
\begin{equation}
 R' \dot{G} - 2 \dot{R} G \ {\nu}' = 0  \label {Dsfe07}
\end{equation}
\begin{equation}
\hspace{.5in} G - H = 1 - \frac{{F}}{R^{N-3}} \; \; .
\label {Dsfe08}
\end{equation}
where $F = F(t, r)$ is an arbitrary function, and has an interpretation of the mass function for the cloud.
 It gives the total mass in a shell of comoving radius $r$ on any spacelike hypersurface $t =$ const. 
The energy conditions impose condition on $F$ that $F(t,r) \geq 0$. 
The physically realistic model preserves the regularity at the initial epoch, so we have $F(t_{i}, 0) = 0$,
 that is, the mass function should vanish at the center of the cloud. Since we are considering collapse, 
we have $\dot{R} < 0$, i.e. the physical radius $R$ of the collapsing cloud keeps decreasing in time 
and ultimately reaches $R = 0$, which denotes the shell-focusing singularity where all matter shells collapse to a zero physical radius. 
The functions $G$ and $H$ are defined as $G(t, r) = e^{-2 \psi} {R'}^2$ and $H(t, r) = e^{-2 \nu} \dot{R}^2$.

We introduce a new function  $v(t, r)$ by
$v(t, r)= R/r$,  
and further use the scaling independence of the comoving
coordinate $r$ to write
\begin{eqnarray}
R(t, r) = r \; v(t, r) \hspace{1in} \nonumber \\
\text{where} \; \;
v(t_{i}, r) = 1 ; \; \; v(t_{s}(r), r) = 0 ; \; \; \dot{v} < 0 \label{Df09}
\end{eqnarray}
The time $t = t_{s}(r)$ corresponds to the shell-focusing singularity at $R = 0$, where all the matter shells collapse to a
vanishing physical radius. The
introduction of the parameter $v$ as above allows us
to distinguish the spacetime singularity from the regular
center, with $v = 1$ at the initial epoch, including the center
$r = 0$, which then decreases monotonically with time
as collapse progresses to value $v = 0$ at the singularity
$R = 0$. At the regular center the mass function $F(t,r)$
behaves suitably so that the density remains finite
and regular there at all times till the occurrence of singular epoch. 
A general mass function for the cloud can be
considered as
\begin{equation}
{F}(t,r) = r^{N-1} {\mathcal{M}} (r,v) \label {Dsm12}
\end{equation}
where ${\mathcal{M}}(r,v)$ is regular and continuously twice
differentiable. Using equation (\ref {Dsm12}) in equation (\ref
{Dsfe05}), we obtain
\begin{equation}
\rho = \frac{N-2} {2} \times  \frac{(N-1) {\mathcal{M}} + r [
{\mathcal{M}}_{, r} + {\mathcal{M}}_{, v}
 \ v']}{ v^{N-2} ( v + r v')} \hspace{.1in}  =
  - \frac{(N-2) {\mathcal{M}}_{, v}} {2 k \ v^{N-2}} \; \; .
 \label {Dsm13}
\end{equation}
Then as $ v \rightarrow 0 \; , \rho \rightarrow \infty $ and $p
\rightarrow \infty $ i.e. both the density and pressure blow up at
the singularity. At the initial epoch, the regular density
distribution takes the form
\begin{equation}
\rho_{o}(r) = \frac{N-2} {2} [r {\mathcal{M}}(r,1)_{,\ r} + (N-1)
{\mathcal{M}}(r,1) ] \; \; . \label {Dsie14}
\end{equation}
We rearrange equation (\ref{Dsm13}) as follows
\begin{equation}
 k r {{\mathcal{M}}}_{,\ r} + W(r,v)
{{\mathcal{M}}}_{,\ v} = - (N-1) k {\mathcal{M}} \label {Dsv14.1}
\end{equation}
where
\begin{equation}
W(r,v) = (k + 1) r v' + v  \; \; . \label {Dsv15}
\end{equation}
Equation (\ref {Dsv14.1}) has many classes of solutions but only
those classes of solutions should be considered which satisfy
energy conditions, which are regular and gives $\rho \rightarrow
\infty$ as $v \rightarrow 0$. This means the energy conditions and
equation of state $ p = k \rho$ isolate the class of functions
${{\mathcal{M}}}(r,v)$ so that the mass function ${F}(t,r)$,
the metric function $\nu (t,r)$ and the function $\textbf{b}(r)$
(to follow) evolve as the collapse begins according to the 
field equations. The energy conditions impose constraint on ${\mathcal{M}}(r,v)$ as
${\mathcal{M}}_{,\; v} < 0$. Hence we set
 \cite{Rpsj}
\begin{equation}
{\mathcal{M}}(0,v) = \frac{{\mathcal{M}}_{o}} {v^{(N-1)k} } \label {Mo1}
\end{equation}
where ${\mathcal{M}}_{o}$ is a positive constant, such that energy density
 $\rho \rightarrow \infty$ as $v \rightarrow 0$.
On integrating equation (\ref {Dsfe06}), we obtain the general
 metric function,
\begin{equation}
\nu(r,v) = - l [  \ln (\rho)  ]. \label {Dcv14}
\end{equation}
where $ l = \frac{k} {k + 1} $.
Define a suitably differentiable function $A(r, v)$ by
 $A(r, v)_{, v} = {\nu'} / {R'}$ so that equation (\ref {Dcv14})
 takes the form \cite{Arg2}
\begin{equation}
A(r,v)_{\; , \; v} = - \frac{l {\rho}'} { {\rho} R'}. \label {Decv15}
\end{equation}
At the initial epoch, we have
$\left[A(r,v)_{\; , \; v} \right]_{v=1} = - {l \ \rho_{o}'(r)}/ 
{ \rho_{o}(r)}  
$
whereas using equation (\ref {Dsm13}), the relation between the
function ${{\mathcal{M}}}$ and $A$, at all epochs is given by
\begin{equation}
A(r,v)_{\; , \; v} \; R' = - l  \left[ \; \ln
\left( - \frac{(N-2) {{{\mathcal{M}}}}_{, v}} {2 k \ v^{N-2}} \right)
\right]'. \label {Dtp17}
\end{equation}
From above, it is clear that ${{\mathcal{M}}}(r,v)$ is a decreasing
function with respect to $v$ and if we consider a smooth initial
profile for the density in such way that density gradient vanishes
at the center then we have $A(r,v) = r g_{o}(r,v)$ where
$g_{o}(r,v)$ is another suitably differentiable function.

Also, the use of definition of $A(r,v)$ in equation (\ref
{Dsfe07}) yields
\begin{equation}
G(t,r) = \textbf{d}_{o}(r) \; e^{2rA} \label {Dg18}
\end{equation}
where $ \textbf{d}_{o}(r) $ is another arbitrary continuously
differentiable function of $r$. Following comparison with the dust
model, we can write
\begin{equation}
\textbf{d}_{o}(r) = 1 + r^2 \; \textbf{b}(r)  \label {Db17}
\end{equation}
where $\textbf{b}(r)$ is the energy distribution function for the
collapsing shells.
 Define a function $\textbf{h}(r,v)$ as
\begin{equation}
 \textbf{h}(r,v) = \frac{e^{2rA} - 1}{r^2}  \label {Dh19}
\end{equation}
and substituting this equation, together with  
(\ref{Dg18}) and (\ref {Db17}) in equation (\ref {Dsfe08}), we get
\begin{equation}
 v^{(N-3)/2} \; \dot{v} = - {\rho}^{-l} \sqrt{v^{N-3}
 \textbf{h}(r,v) + \textbf{b} v^{N-3} e^{2rA}
  + {{\mathcal{M}}}} \;  \label {Dsv20}
\end{equation}
where negative sign is chosen since, for the collapse, $ \dot{v} <
0 $. Integrating the above equation , we have
\begin{equation}
 t(v,r) = \int_{v}^{1}  \frac{ v^{(N-3)/2} dv}{{\rho}^{-l}
 \sqrt{v^{N-3} \textbf{h}(r,v) + \textbf{b} v^{N-3} e^{2rA}
  + {{\mathcal{M}}}}} \; \; .
\hspace{.4in} \label {Dsi21}
\end{equation}
In above equation, the variable $r$ is treated as a constant.
Expanding $t(v,r)$ around the center, we get
\begin{equation}
t(v,r) = t(v,0) + r \chi(v) + {\mathcal{O}}(r^2) \label {Dt22}
\end{equation}
where the function
\begin{equation}
\chi(v) = \frac{dt}{dr} \Big{|}_{r=0} = - \frac{1}{2} \int_{v}^{1}
 \frac{v^{(N-3)/2} {\mathcal{B}}_{,r}(0,v) } {{\mathcal{B}}(0,v)^{3/2}} dv
 \label {Dch23}
\end{equation}
and
\begin{equation}
{\mathcal{B}}(r,v) = {\rho}^{-2 l} \left[v^{N-3}
\textbf{h}(r,v) + \textbf{b} v^{N-3} e^{2rA} + {{\mathcal{M}}}(r,v)\right] \;
\;  . \label {Dchh24}
\end{equation}
The time when the central singularity develops is given by
\begin{equation}
t_{s_{o}} = \int_{0}^{1}  \frac{ v^{(N-3)/2} dv}
{{\mathcal{B}}(0,v)^{1/2}} \; \; . \label {Dth25}
\end{equation}
The time for other collapsing shells to reach the singularity can
be expressed by
\begin{equation}
t_{s}(r) \equiv t(0,r) = t_{s_{o}} + r \chi(0) + {\mathcal{O}}(r^2) \; \; .
\label {Dt26}
\end{equation}
\begin{equation}
\text{where} \; \; \chi(0) = \lim_{v \rightarrow 0} \chi(v). \label {Dscho25}
\end{equation}
Now, it is clearly seen that the value of $\chi(0)$ depends on the
free functions $\textbf{b}(0),{{\mathcal{M}}}(0,v)$ and $
\textbf{h}(0,v)$, which in turn, depend on the initial data at the
initial surface $ t=t_{i}$. Thus, a tangent to the singularity
curve $ t = t_{s}(r) $ is completely determined by the given set
of density, pressure, velocity function $\nu$ and function
$\textbf{b}$. Further, from equation (\ref {Dsv20}), on $v=$ const. surface, we can write
\begin{equation}
v^{\frac{(N-3)}{2}} \;  v' = \chi(v) \; {\mathcal{B}}(0,v)^{1/2} + {\mathcal{O}}(r^2) \; \; .
\label {Dst24}
\end{equation}

Let us analyze the nature of $\chi(0)$ from the point of view of equation (\ref{Dch23}). 
The existence of the limit depends on the cumulative effect of all
the terms present in the equation but since $
{{\mathcal{B}}(0,v) }> 0$ as $v \rightarrow 0$, provided
${{\mathcal{M}}}(0,0) > 0$ and $v \in [0,1]$, therefore positive or
negative sign of $\chi(0)$ absolutely depends on the sign of
 ${{\mathcal{B}}_{, \; r}(0,v) }$.
 
 Now, $ {{\mathcal{B}}(0,v) }> 0$ and
if
\begin{eqnarray}
{{\mathcal{B}}_{, r}(0,v) } = - \frac{2  l \rho'(0,t)
{{\mathcal{B}}(0,v) }} { \rho}  + \frac{ \left[ v^{N-3} ( h_{, r}(0,v)
+ \textbf{b}'(0) ) + {{\mathcal{M}}}_{,  r}(0,v) \right]} {\rho^{2 l}} < 0  \label{BR01}
\end{eqnarray}
 then integrand in equation (\ref{Dch23} ) is positive and hence, $\chi(0) >
 0$. This is possible, since density is very high in the
 neighbourhood of the singularity and it decreases as $r$
 increases from $r=0$, therefore, $\rho'(0,t) $ should be negative
to be physically reasonable and hence sign of ${{\mathcal{B}}_{,
\; r}(0,v)  }$ depends on the sign of $\left[ v^{N-3} [ \textbf{h}_{, \;
r}(0,v) + \textbf{b}'(0) ] + {{\mathcal{M}}}_{, \; r}(0,v) \right]$ i.e.
depends solely on the initial data. The introduction of an equation of state has ensured
that the system of Einstein equations is closed and so freedom to specify any function is exhausted. Therefore, the functions $\textbf{h}(r,v)$  be written in terms of $\textbf{b}(r)$ and ${\mathcal{M}}(r,v)$. Hence, the choice of class of
functions $\textbf{b}(r)$ and ${\mathcal{M}}(r,v)$ shall decide the final fate of
gravitational collapse. So, the occurrence of NS or BH
phases exist, depending on the choice of the class of function
${\mathcal{M}}(r,v)$.

\section {Radial null geodesics}
Now, we investigate below when there will be families of null geodesics
emanating, which will be future directed and outgoing, and which terminate in the past at the singularity, thus making
the communication from the singularity to an outside observer possible, as opposed to a black hole situation. 
Hence, for examination of the nature of
central singularity at $ R = 0, r = 0$, we consider the equation of outgoing radial null geodesics,  given by,
\[ \frac{dt} {dr} = e^{\psi - \nu}. \]
  Further, we write the null geodesic
equation in terms of the variables ( $ u= r^{\beta}, R $),
choosing $ \beta = \frac{1}{1-k}[\frac{(N+1)}{(N-1)} -k ]$, and using equation (\ref {Dsfe08}), we
have 
\begin{equation}
  \frac{dR}{du} = \frac{1}{\beta} \left( \frac{R}{u} +
\frac{ v'v^{\frac{(N-3) -(N-1)k}{2}} } { (\frac{R}{u})^{\frac{[(N-3) -(N-1)k]}{2}} } \right) \left( \frac{1 -
\frac{F}{R^{N-3}} } { \sqrt{G}[ \sqrt{G} + \sqrt{H}]} \right)  \; . \label
{Nrn25}
\end{equation}
If the null geodesics terminate at the singularity in the past
with a definite tangent, then at the singularity, we have 
$dR/du > 0 $, in the $(u, R)$ plane with a finite value.
All other points $r> 0$ on the singularity curve are covered
since $ F/R^{N-3} \rightarrow \infty $ with $  dR/du \rightarrow
{-\infty}$ and only the singularity at the centre $ r=0$ could be
naked. For the case, when $R' > 0$ near the central singularity,
we have
\begin{equation}
x_{o} = \lim_{t\rightarrow t_{s}} \lim_{ r\rightarrow {o}}
\frac{R}{u} =  \frac{dR}{du} \Big{|} _{t\rightarrow
t_{s},r \rightarrow {o}} \label {pr26}
\end{equation}
and with the use of equation (\ref {Dst24}) for $v'$,  equation 
(\ref {Nrn25})  yield  
 \begin{eqnarray}
  {x_{o}}^{\frac{(N-1)(1-k)}{2}} = \frac{(1-k)(N-1)}{2} \sqrt{{\mathcal{M}}_{o}} 
 \left(\frac{2} {{\mathcal{M}}_{o} (N-1) (N-2)} \right)^{\frac{k} {k+1} }  \chi(0) \label{Chi01}
 \end{eqnarray}
 and the radial null geodesic emerging from the
singularity in $(R,u)$ co-ordinates is $ R = x_{o} u$, or in
$(t,r)$ plane, it is given by 
\[ t - t_{s}(0) = x_{o} \; r^{\beta} \; .\]
Therefore, $ x_{o} > 0$ iff $ \chi(0) > 0 $, and hence $\chi(0) > 0$ is a sufficient condition for the occurrence of the NS at the center of the cloud as the end state of gravitational collapse of a sufficiently dense star when it loses its equilibrium state, and continual collapse begins with regular initial data of density and pressure profiles
$p = k \rho$. There will be 
radially future outgoing null geodesics emanating from the singularity,
giving rise to a locally NS at the center. However, if $ \chi(0) < 0 $ then we have a black hole solution, as there will be no
such trajectories coming out. Also, it is observed from equation (\ref{Chi01}) that parameter $k$ 
( because $(1-k)>0$) and dimensionality $N$ do not take part in the decision making of the sign of $\chi(0)$. 
Thus the final fate of collapse of a massive star leads to either a 
BH or a NS with appropriate choice of initial data for $N \geq 4$. For $ \chi(0) = 0 $ , we will have
to take into account the next higher order non-zero term in the
singularity curve equation, and a similar analysis can be carried
out by choosing a different value of $\beta$.

The radially emitted rays from central singularity may fall back to the singularity
under the intense pull of gravity. So, for the central singularity to be at
least locally naked, it should satisfy the necessary condition due to the
apparent horizon curve given by equation $F/R^{N-3}=1$
which gives the boundary of the 
trapped surface region of the spacetime. 

\section{Initial data relation through b(r) leads to NS}
In this section, we prove the assertion mentioned in the
introduction. We choose $b(r)$ to satisfy the differential
equation on a constant $v$-surface
\begin{equation}
 \frac{1}{2} \frac{v^{(N-3)/2} {\mathcal{B}}_{,r}(r,v) } {{\mathcal{B}}(r,v)^{3/2}} = \texttt{C}(r,v)
 \label {br1}
\end{equation}
for $ \ 0 \leq r \leq r_{b} \ $ where  $ \texttt{C}(r,v) $ is a continuous
function defined on $ {\mathcal{D}}$ such that $ \texttt{C}(0,v) < 0$
for all $ v $ in $[0,1]$ and ${\mathcal{B}}(r,v)$ is given by equation
(\ref{Dchh24}).
 It follows that
\begin{equation}
\chi(0) = \lim_{v \rightarrow {0}} \chi(v) = - \int_{0}^{1} \texttt{C}(0,v)
dv > 0 . \label {nc29}
\end{equation}
This condition ensures that central shell-focusing singularity
will be naked.

We, now, discuss the existence of $\textbf{b}(r)$ as a solution of the
differential equation (\ref {br1}). We put
\begin{equation}
y(r,v) = 
 {\rho}^{-2 \; l} \left[v^{N-3}
\textbf{h}(r,v) + \textbf{b} v^{N-3} ( 1 + r^2 \textbf{h}(r,v)) + {{\mathcal{M}}}(r,v)\right] 
    \label {br3}
\end{equation}
a continuous function of $r$, in equation (\ref {br1}) and
rearranging it, we get
\begin{equation}
\frac{dy}{dr} = \frac{1}{v^{(N-3)/2}} \left[ 2 \; \texttt{C}(r,v) y^{3/2}
  \right] \equiv z(y,r) \label {br4}
\end{equation}
with the initial condition
\begin{equation}
 y(0,v) = {\rho(0,v)}^{-2 \; l} \left[v^{N-3}
\textbf{h}(0,v) + \textbf{b}(0) v^{N-3} + {{\mathcal{M}}}(0,v)\right]   \label {br5}.
\end{equation}
Let us ensure the existence of $ C^{1} $ - function $y(r,v)$ as a
solution of above initial value problem defined throughout the
cloud. The function $z(y,r)$ is continuous in $r$, with $y$
restricted to a bounded domain. With such domain of $r$ and $y$,
$z(y,r)$ is also  $ C^{1} $ - function in $y$ which means $z(y,r)$
is Lipschitz continuous in $y$. Therefore, the differential
equation (\ref {br4}) has a unique solution satisfying initial
condition (\ref {br5})
\cite{Isbsrvs}, \cite{JMS}.

  Further, we can ensure that the solution will be defined over the
entire cloud i.e. for all $r$ in $ [0,r_{b}]$ by using the freedom
in the choice of arbitrary function $\texttt{C}(r,v)$. For this , we
consider  the domain $[0,r_{b}] \times [0,d]$ for some finite $d$.
Let $S_{p} = \sup |z(y,r)|$. Then the
differential equation (\ref {br4}) has a unique solution defined
over the entire cloud provided \cite{Ave}
\begin{equation}
r_{b} \leq \inf (r_{b}, \frac{d}{S_{p}}) = \frac{d}{S_{p}}  \label {ec33}.
\end{equation}
This yields
\begin{equation}
\max_{{0\leq r \leq r_{b}},   {0 \leq y \leq d} } \hspace{.1in}
\left| \frac{1}{v^{(N-3)/2}} \left[ 2 \texttt{C}(r,v) y^{3/2}
  \right] \right|
\leq \frac{d}{r_{b}} \label {ma34}.
\end{equation}
Condition (\ref {ec33}), in turn, will be satisfied if the weaker
condition
\begin{equation}
0 \leq |\texttt{C}(r,v)| \ y^{3/2} \leq \frac{d \; v^{(N-3)/2} } {2 \; r_{b}}
\label {sm35}
\end{equation}
holds for all $r$ in $[0,r_{b}]$.

The collapsing cloud may start with  $r_{b}$ small enough so that
the expression 
\ $ {d \; v^{(N-3)/2} }/{2 \; r_{b}}$ \
which is always positive, satisfies the condition (\ref
{sm35}) with $y$ restricted to a bounded domain. We then have
infinitely many choices for the function
$\texttt{C}(r,v)$, which is continuous and satisfies conditions  
(\ref {sm35}) and $\texttt{C}(0,v) < 0$ for each choice of $v$. For each such
$\texttt{C}(r,v)$, there will be a unique solution $y(r,v)$ of the
differential equation (\ref {br4}), satisfying initial condition
(\ref {br5}), defined over the entire cloud and in turn, there
exist a unique function $\textbf{b}(r)$ for each such choice of $\texttt{C}(r,v)$,
that is given by the expression
\begin{eqnarray}
\textbf{b}(r) = \frac{1} { v^{N-3} [ 1 + r^2 \textbf{h}(r,v)] } 
 \Big\{ y(r,v) \left( \frac{N-2}{2} \right)^{2 \; l} \times \hspace{1in} \nonumber \\
\left( \frac{(N-1){\mathcal{M}} + r ({\mathcal{M}}_{,r} + {\mathcal{M}}_{,v} v') } { v^{N-2} (v + r v') } \right)^{2 \; l}    
 - v^{N-3} \textbf{h}(r,v) -  {{\mathcal{M}}}(r,v)   \Big\} 
     \label {br9}
\end{eqnarray}
over $[0,r_{b}]$. Thus, for a given constant $v$-surface and given
initial data of mass function $ F(t,r) = r^{N-1}{{\mathcal{M}}}(r,v) $
and $ \textbf{h}(r,v) = (e^{r^2 g_{2}(r,v)} - 1) /{r^2} $ satisfying
physically reasonable conditions (expressed on ${{\mathcal{M}}}$),
there exists infinitely many choices for the function $\textbf{b}(r)$   \
such that condition $\texttt{C}(0,v) < 0 $ is satisfied. The condition
continues to hold as $v \rightarrow 0$, because of continuity.
Hence, the central singularity developed in the collapse is a
naked singularity established using initial data relation through 
$\textbf{b}(r)$ given in equation (\ref{br9}).
\section {Stability of Naked Singularity}
The analysis in sections 2 and 4 shows that the only conditions on the
initial data which evolve the collapse into a naked singularity
are the energy conditions and $\texttt{C}(0,v) < 0$. Hence, for stability, we 
examine these conditions only. What we now show is that, the set
of such initial data satisfying these conditions forms an open
subset of the space of all initial data. For this, we use the
technique of Saraykar and Ghate \cite{Asg}, \cite{Isbsrvs}. 

We first consider energy conditions 
which are given by
${{\rho}{\geq}}0$ and ${\rho+p{\geq}} 0$ but since $p = k \rho$,
${{\rho}{\geq}}0$ is the only sufficient condition required in this set up and 
can be written in the form
\begin{equation}
\left[ (N-1) {{\mathcal{M}}} + r {{\mathcal{M}}}_{,r} + r v'
{{\mathcal{M}}}_{,v}\right] \equiv E \geq 0, \label{eq:E01}
\end{equation}
on the domain ${\mathcal{D}}$.
 Here, we have used an assumption
that $v$ is an increasing function of $r$ so that $v'\geq 0$ on
any surface $ t = t_{j}$ say, so that no shell-crossing
singularity condition $(ie. R' > 0)$ holds. So, this is a valid
assumption. 

We assume that $X$ be an infinite dimensional Banach space of all
$C^{1}$ real-valued functions defined on ${\mathcal{D}}$, endowed
with the norms
\begin{eqnarray*}
\parallel {{\mathcal{M}}}(r,v) \parallel = \sup_{{\mathcal{D}} } |{{\mathcal{M}}}  |
+ \sup_{{\mathcal{D}} } |{{\mathcal{M}}}_{,r}| + \sup_{{\mathcal{D}} }
|{{\mathcal{M}}}_{,v}|    \hspace{.8in} \nonumber 
\end{eqnarray*}
These norms are equivalent to the standard $C^{1}$ norm
\begin{eqnarray}
\parallel {{\mathcal{M}}}(r,v) \parallel = \sup_{{\mathcal{D}} } (|{{\mathcal{M}}} |
+ |{{\mathcal{M}}}_{,r}| + |{{\mathcal{M}}} _{,v}|) \hspace{.6in} \
 \label {def37}
\end{eqnarray}
Let ${\mathcal{G}}_{1} = \{ {{\mathcal{M}}}(r,v):{{\mathcal{M}}}> 0, {{\mathcal{M}}}
$ \ is $ \ C^{1},  \ E > 0 $ \ on $
{\mathcal{D}} \}$ \ be a subset of $X$.

We show that ${\mathcal{G}}_{1}$ is an open subset of $X$. For
${{\mathcal{M}}}$ in ${\mathcal{G}}_{1}$, let us put $\delta =
\min({{\mathcal{M}}})$, $\gamma = \min( E)$, 
and $\lambda_{3} = \max(v')$ for varying $r$ in
$[0,r_{b}]$ and $ v \; \epsilon [0,1]$, the functions involved herein
are all continuous functions defined on a compact domain
${\mathcal{D}}$ and hence, their maxima and minima exist.
 We define a positive real number
 \begin{eqnarray*}
  \mu = \frac{1}{2} \min \Big\{\delta , \frac{\gamma}{3 (N-1)},
\frac{\gamma}{3 r_{b}}, \frac{\gamma}{3 r_{b} \lambda_{3}} \Big\}.
\end{eqnarray*}
Let ${{\mathcal{M}}}_{1}(r,v)$ \ be  \ $C^{1}$ \ in \ $ {\mathcal{D}}$
\ with \ $\parallel {{\mathcal{M}}} - {{\mathcal{M}}}_{1}
\parallel  < \mu$.
Using definition (\ref {def37}), we get $| {{\mathcal{M}}}_{1} -
{{\mathcal{M}}} | < \mu$, $| {{\mathcal{M}}}_{1,r} - {{\mathcal{M}}}_{,r} | <
\mu$  \ and $| {{\mathcal{M}}}_{1,v} - {{\mathcal{M}}}_{,v} |< \mu$ over ${\mathcal{D}}
$. Therefore, for choice of $\mu$, the respective inequalities are
\begin{eqnarray}
{{\mathcal{M}}}_{1} > {{\mathcal{M}}} - \frac{\delta}{2} > 0 ,  \label{IEQ01} \\
(N-1) |{{\mathcal{M}}}_{1} - {{\mathcal{M}}} | < \frac{\gamma}{6} ,
\label{IEQ02} \\
 r |{{\mathcal{M}}}_{1,r} - {{\mathcal{M}}}_{,r} | \leq r_{b}
|{{\mathcal{M}}}_{1,r} - {{\mathcal{M}}}_{,r} | < \frac{\gamma}{6} ,
 \label{IEQ03} \\
 r v' | {{\mathcal{M}}}_{1,v} - {{\mathcal{M}}}_{,v} |  \leq r_{b} \lambda_{3}
 | {{\mathcal{M}}}_{1,v} - {{\mathcal{M}}}_{,v} | < \frac{\gamma}{6}, \label{IEQ04}
 \label {de38}
 \end{eqnarray}
 that are satisfied on ${\mathcal{D}}$.
 Combining inequations (\ref{IEQ02},\ref{IEQ03},\ref{IEQ04}) from above,
 we have
\[  (N-1) |{{\mathcal{M}}}_{1} - {{\mathcal{M}}} | + r
|{{\mathcal{M}}}_{1,r} - {{\mathcal{M}}}_{,r} | + r v'|{{\mathcal{M}}}_{1,v} -
{{\mathcal{M}}}_{,v} |< \frac{\gamma}{2} < {\gamma}  \leq E
\]
Using equation (\ref{eq:E01}), we can write 
\[ |\; [(N-1) {{\mathcal{M}}}_{1} + r
{{\mathcal{M}}}_{1,r}+ r v' {{\mathcal{M}}}_{1,v}] - E \; | < E 
 \; \text{where} \;  E > 0 \; \text{on} \; {\mathcal{D}}. \]
Hence, \ $ [(N-1) {{\mathcal{M}}}_{1} + r {{\mathcal{M}}}_{1,r}+ r v' {{\mathcal{M}}}_{1,v}] > 0 $ on $
{\mathcal{D}}$. 
\\
Thus,\ $ {{\mathcal{M}}}_{1}  > 0$,
${{\mathcal{M}}}_{1}$ is $C^{1}$ and $ [(N-1) {{\mathcal{M}}}_{1} + r
{{\mathcal{M}}}_{1,r}+ r v' {{\mathcal{M}}}_{1,v}] > 0 $ throughout ${\mathcal{D}}$.
 Therefore, ${{\mathcal{M}}}_{1}(r,v)$ also
lies in ${\mathcal{G}}_{1}$ \ and hence,\ ${\mathcal{G}}_{1}$ is an open
subset of $X$.
\\
Let  ${\mathcal{G}}_{2} = \{ {{\mathcal{M}}}(r,v):\texttt{C}(0,v) < 0$, \
defined on ${\mathcal{D}} \}$ \ be a subset of \ $X$.
\\
Using the same technique as above, we can prove that the set ${\mathcal{G}}_{2}$ is an open set in $X$. Therefore, ${\mathcal{G}}= {\mathcal{G}}_{1} \cap {\mathcal{G}}_{2}$ 
is an open set in $X$.
\\
Similarly, the subset  ${\mathcal{H}}$ of $X$ defined by 
\[{\mathcal{H}} = \{ \textbf{h}(r,v):\texttt{C}(0,v) < 0, \;
\text{defined on} \; {\mathcal{D}} \}\] 
 would form an open subset in $X$. Hence,
\[ {\mathcal{G}} \times {\mathcal{H}} = 
\{ ( {{\mathcal{M}}}(r,v), \textbf{h}(r,v) ):{{\mathcal{M}}}> 0, {{\mathcal{M}}}
 \; \text{is} \; C^{1},  \; E > 0 \; \text{and} \; \texttt{C}(0,v) < 0 \} \]
is an open subset in the product space $X \times X$.
\\ 
 Taking $({\mathcal{M}}_{1}, \textbf{h}_{1}(r,v))$ in the neighbourhood of
$({\mathcal{M}},\textbf{h}(r,v))$ in $G \times {\mathcal{H}}$, and using
equation (\ref {sm35}) analogously for $ {\mathcal{M}}_{1}$ and
$\textbf{h}_{1}$, we have a choice of infinitely many $\texttt{C}_{1}(r,v)$, such
that for each such $\texttt{C}_{1}(r,v)$, there will exist a unique
$\textbf{b}_{1}(r)$ so that the initial data of mass function $r^{3}
{\mathcal{M}}_{1}$ and $\textbf{h}_{1}(r,v)$ together will lead the collapse
to formation of a naked singularity. Thus, naked singularity
arising from $({\mathcal{M}},\textbf{h}(r,v))$ is $C^{1}$ - stable in the
sense defined above but since in earlier analysis the initial data functions 
are at least $C^{2}$, therefore NS arising from initial data is $C^{2}$-stable.

The analysis given above, guarantee the existence of a metric
function $ \nu(t,r)$ for a given initial data set. Such choice of
$\nu(t,r)$ and expressions for $G$ and $H$ together will yield the
metric (\ref {Dm01}) as an exact solution leading to the occurrence
of naked singularity. 

Reverting the sign of $\texttt{C}(0,v)$ i.e. considering $\texttt{C}(0,v) > 0$,
we can similarly prove that the occurrence of black hole is also $C^{2}$-stable.
\subsection {Genericity of Black Holes and Naked Singularities}
We have seen that in Type I matter field collapse, the end state of collapse is governed by the choice of initial data.
 We have also shown that the set of initial data leading the collapse to a naked singularity 
or a black hole forms an open subset of the full set of initial data, and thus both the occurrences are stable.

It is also clear that these sets as subsets of initial data set leading the collapse to BH and NS are mutually disjoint. 
Since both of them are open, none of them  can be dense in $X \times X$. 
Thus, in the strict Mathematical sense \cite{Irajm,Isbsrvs}, both these occurrences are not generic.

Nevertheless, each of these sets are substantially big sets. Following the measure theoretic analysis 
on infinite dimensional spaces given in Section IV of \cite{JMS}, it can be argued that these sets have a non-zero measure.
In this sense, we can say that these occurrences are `generic',
 namely the set of initial data in each case is open and has a non-zero measure. 
\section{Conclusions}

Our main conclusions are as follows: 

\noindent 1(a). Given
a $C^{2}$- mass function $ {\mathcal{M}}(r,v)$ and a $C^{1}$-
function $\textbf{h}(r,v)$, on any $v =$ const. surface, we find a $C^{1}$-
energy distribution function $\textbf{b}(r)$ such that the collapse ends in
a naked singularity. 

\noindent 1(b). With physically reasonable conditions put
on mass function ${\mathcal{M}}$, and with $C^{1}$- function
$\textbf{h}(r,v)$, the initial data consisting of $({\mathcal{M}},\textbf{h}(r,v))$
leading the collapse to formation of a naked singularity, forms an open subset
of the space $X \times X$. This establishes stability of naked
singularity with respect to initial data.

\noindent 2. Similar analysis for the case $\chi(0)< 0$ ( that is $\texttt{C}(0,v)>
0$ for all $v$ in $[0,1]$) shows that the occurrence of black hole is also stable. 
Thus, both the occurrences, namely, black holes and naked singularities are stable. 
Further, we state that these occurrences are {\it generic},
 in the sense that the set of initial data in each case is open and has a non-zero measure. 

\noindent 3. Also, it is found that equation of state parameter $k$ 
and dimensionality $N$ of spacetime do not take part in the decision making
of the sign of $\chi(0)$. So for a critical value $\chi(0) > 0$, a spectrum of radial null geodesics can emanate 
from the central singularity for $k \in [0,1]$ and for $N \geq 4$ but many of them may fall back to the singularity.

In our analysis, in equation  (\ref{nc29}), the expression of $\chi(0)$ is expressed in terms of smooth functions which are general in nature. 
 Therefore, the formation of NS/BH depends on the choice of initial data of these functions. 
 Furthermore, in reference \cite{Rpsj}, the equation (74) of apparent horizon curve 
whose increasing property is a necessary condition for the existence of a locally naked singularity,
 is specified in terms of general functions $A(r,v), \nu(t,r), h(r,v), b(r), {\mathcal{M}}(r,v)$. 
Therefore outcome of NS depends on these functions. 
 Hence, we cannot say definitely that increasing $N$ would always lead to BH.

In the analysis of dust collapse, apparent horizon is a well behaved surface
 and hence one can draw a definite conclusion for higher values of $N$ 
but since our initial data functions $b(r)$ and ${\mathcal{M}}(r,v)$ are general in nature,
 both the outcomes are possible with the appropriate choice of these functions. 
Therefore, our results are independent and general in nature as compared 
 to the results obtained in references \cite{{Drgpj-1},  {rgpj-1}} as these results are subject to the specific choice of initial data.

Further, it can be investigated whether $k$ and $N$ assist in making a locally NS 
to be a globally NS which can be seen by a faraway observer by finding a solution
 to a differential equation (\ref{Dsv14.1}) in terms of ${\mathcal{M}} $ or by choosing a appropriate Taylor series
realistic expansion of ${\mathcal{M}}$ for a given mass function $F(t,r)$.
\section*{Acknowledgement}
Sanjay Sarwe acknowledges the facilities extended by IUCAA, Pune, 
while part of this work was being completed.

\end{document}